\documentclass{vgtc}                          

\ifpdf
  \pdfoutput=1\relax                   
  \usepackage{graphicx}                
  \DeclareGraphicsExtensions{.pdf,.png,.jpg,.jpeg} 
\else
  \usepackage{graphicx}                
  \DeclareGraphicsExtensions{.eps}     
\fi%

\graphicspath{{figures/}{pictures/}{images/}{./}} 

\usepackage{microtype}                 
\PassOptionsToPackage{warn}{textcomp}  
\usepackage{textcomp}                  
\usepackage{mathptmx}                  
\usepackage{times}                     
\usepackage{cite}                      
\usepackage{tabu}                      
\usepackage{booktabs}                  

\onlineid{0}

\vgtccategory{Research}

\vgtcinsertpkg

\title{EduVis: Workshop on Visualization Education, Literacy, and Activities}

\author{Mandy Keck\thanks{e-mail: mandy.keck@fh-hagenberg.at}\\ %
       \parbox{1.5in}{\scriptsize \centering University of Applied Sciences\\ Upper Austria}%
\and Samuel Huron\thanks{e-mail: samuel.huron@cybunk.co}\\ %
     \parbox{1.5in}{\scriptsize \centering Telecom Paris,\\ Institut Polytechnique de Paris\\} %
\and Georgia Panagiotidou\thanks{e-mail: g.panagiotidou@ucl.ac.uk}\\ %
     \parbox{1.3in}{\scriptsize \centering UCLIC,\\ University College London\\}
\and Christina Stoiber\thanks{e-mail: christina.stoiber@fhstp.ac.at}\\ %
     \parbox{1.1in}{\scriptsize \centering St. Pölten University\\of Applied Sciences\newline}
\and Fateme Rajabiyazdi\thanks{e-mail: fatemerajabiyazdi@cunet.carleton.ca }\\ %
     \parbox{1.5in}{\scriptsize \centering Carleton University}
 \and Charles Perin \thanks{e-mail: cperin@uvic.ca}\\ %
     \parbox{1.5in}{\scriptsize \centering University of Victoria}
\and Jonathan Roberts \thanks{e-mail: j.c.roberts@bangor.ac.uk}\\ %
     \parbox{1.3in}{\scriptsize \centering Bangor University}
\and Benjamin Bach \thanks{e-mail: bbach@ed.ac.uk}\\ %
     \parbox{1.2in}{\scriptsize \centering University of Edinburgh}}     

\abstract{This workshop focuses on visualization education, literacy, and activities. It aims to streamline previous efforts and initiatives of the visualization community to provide a format for education and engagement practices in visualization. It intends to bring together junior and senior scholars to share research and experience and to discuss novel activities, teaching methods, and research challenges. The workshop aims to serve as a platform for interdisciplinary researchers within and beyond the visualization community such as education, learning analytics, science communication, psychology, or people from adjacent fields such as data science, AI, and HCI. It will include presentations of research papers and practical reports, as well as hands-on activities. In addition, the workshop will allow participants to discuss challenges they face in data visualization education and sketch a research agenda of visualization education, literacy, and activities.} 

\begin{document}


\maketitle

\section{Motivation}
With the increasing democratization of visualization and the maturing of it as a scientific discipline, we need to promote critical reflection, responsive education, and deliberate engagement with topics in visualization. Data Visualization literacy is the ability to read, write, and create information visualizations \cite{bach2021special} and is crucial for a range of contemporary key skills such as data analysis, quality control, human-AI interaction, effective communication, or education in the general sense. Numerous studies are warning about the false use of visualization \cite{Lo2022, McNutt2020, Cairo2019}, and we have guidelines to inform visualization design \cite{VisGuides2022} but little on how to teach visualization design to novices.
At the same time, we research novel visualization techniques but do little to train our intended users. We produce powerful analysis and editing tools but know little if and how people struggle in the wild using them. Many of us teach topics in visualization, but we have no foundation in how to teach visualization topics, engage with learners, or assess success in learning visualizations. We frequently engage with collaborators we want to empower through our research. Still, we have only vague personal experiences communicating our contributions and engaging our collaborators in co-design. We want more students, junior faculties, and researchers to grow our community, but we do not support those new to our community in practices and education. The broad range of topics (perception, interaction, visualization techniques, design, storytelling, etc.), audiences (children, students, working professionals, designers, data scientists, journalists, educators, decision-makers, etc.), and scenarios (classroom learning, informal learning, collaboration, museums, science fairs, workshops, hybrid and online teaching, scientific collaborations, consulting, etc.) call for a structured and informed approach to visualization knowledge, education, and practices to address questions about effective engagement activities and education resources, learning goals, audiences, diversity, methods, assessment methods and digital support tools.

As one prominent way to engage with visualizations, visualization activities~\cite{huron2020ieee} will be an integral part of our workshop. For example, by engaging with visualization artifacts such as reading, critiquing, explaining, and recreating, or by engaging with the visualization process, such as designing, crafting, conceptualizing. Data visualization activities aim to increase data visualization literacy and can be used in an educational setup but also for co-design, activism, or self-reflection. We see an increase in visualization activities and the need for collecting, disseminating, and evaluating these activities. Examples include the five design-sheet methodology for visualization design~\cite{roberts2015sketching},  workshops for dashboards, data comics and storytelling~\cite{bach2023, bach2018design}, methods for design immersion~\cite{Hall}, design spaces and patterns to practice visualization design~\cite{keck2021}, tangible visualization with physical tokens~\cite{VisKit}, sketching exercises for visualizing two quantities~\cite{Ortiz_2012}, methods for eliciting self-reflection on personal data~\cite{Thudt:2018:SPP:3173574.3173728}, critical thinking sheets~\cite{CriticalThinkingSheets2020}, engaging people with physicalization~\cite{huron2017let}, and activities to improve data literacy~\cite{bhargava2015designing,he2016v}. 

\section{Previous related scientific initiatives}
The visualization community has some track record in discussing education, initiated by a wide range of people and often independent from each other.
Initiatives include 
\begin{itemize}
    \item IEEE VIS workshops: two workshops (2020, 2021) on Data Vis Activities to facilitate learning, reflecting, discussing \cite{huron2020ieee, huron2021ieee} and two workshops (2016, 2017) on Pedagogy of Data Visualization \cite{WorkshopPedagogy2016, WorkshopPedagogy2017}
    \item Dagstuhl seminar (2022) on ``Visualization Empowerment: How to Teach and Learn Data Visualization'' \cite{bach_et_al:DagRep.12.6.83}
    \item Special issue on Visualization Education at IEEE Computer Graphics and Application (2021)\cite{bach2021special}
    \item IEEE VIS 2015 Panel: Vis, the next generation: Teaching across the researcher-practitioner gap \cite{Panel2015}
\end{itemize}    

These previous events have demonstrated that education is a topic of permanent interest to the visualization community. This interest is also reflected in the list of visualization researchers and educators on our program committee who would like to support a workshop on visualization education, literacy, and activities at IEEE VIS (see section \ref{pccommittee}).\newline
Each event yielded high-quality outputs (papers, books, activities, etc.) and contributed to community building. It is thus timely to streamline these efforts and provide a platform for education and engagement practices in visualization and we hope to eventually be able to establish a permanent forum at IEEE VIS, akin to other communities such as EduCHI (ACM SIGCHI) \cite{EduCHI2023}, the education track at Eurographics\cite{EduEurogaphicsConference2021}, or the public resources and activities created by the Data-Viz Society. 
This forum would benefit students, junior faculty members new to teaching, senior researchers in planning outreach and education, as well as researchers and practitioners outside the (academic) visualization community. 

The Dagstuhl seminar 2022 on ``Visualization Empowerment: How to Teach and Learn Data Visualization'' \cite{bach_et_al:DagRep.12.6.83} and the last IEEE VIS workshop on Data Vis Activities \cite{huron2021ieee} established a set of open questions and challenges that need to be addressed as a community. Hence, we plan to organize specific work groups around these most pressing issues (see section \ref{workshop_activities}). Moreover, participants of these events seemed to converge around the need for a reporting platform to capture the variety of activities available. We thus plan on setting up a barebones version of such a platform for this workshop to continue working on it together with this year's participants.

\section{Workshop Goals}
A workshop on Visualization Education, Literacy, and Activities can bring new people to IEEE VIS who are outside the traditional visualization network. This includes people from education, learning analytics, science communication, psychology, or people from adjacent fields such as data science, AI, and HCI. Another goal of this workshop is to bring together newly appointed and experienced faculty to share experiences and discuss novel datavis activities, teaching methods, and challenges. Furthermore, this format is intended to serve as a platform for researchers to foster interdisciplinary exchange and share research results and best practices.\\

With this workshop, we want to achieve the following goals:
\begin{itemize}
    \setlength{\itemsep}{2pt}
    \setlength{\parskip}{0pt}
    \setlength{\parsep}{0pt}
\item build a permanent forum and interdisciplinary community around teaching data visualization, open to researchers, students, and practitioners outside the traditional VIS community
\item publish research on visualization literacy, education and teaching, activities, and practices, etc. (see section \ref{topics} for a more detailed list of topics)
\item discuss best practices to teach data visualization to diverse audiences (e.g., children/adult learning, data journalists/data scientists/computer scientists/designers) and in different scenarios (onsite, online, hybrid)
\item share visualization educational tools, material, and processes
\item collect, test, and systematize learning activities
\item discuss higher-level issues concerning human-centered approaches to visualization, visualization design, and education
\end{itemize}

\section{Scope of Topics}
\label{topics}
The workshop topics include, but are not limited to:
\begin{itemize}
    \setlength{\itemsep}{2pt}
    \setlength{\parskip}{0pt}
    \setlength{\parsep}{0pt}
\item Visualization literacy
\item Learning goals and learning methods
\item Evaluation methods and learning analytics
\item Educational tools
\item Visualization activities
\item Pedagogy in visualization
\item Hybrid and online teaching
\item Reflective and research practices
\item Understanding audiences
\item Guidelines, strategies, and guidance for education
\item Debate and discussions on visualization guidelines and well-established knowledge
\item Knowledge dissemination
\item Challenges and personal experiences
\item Informal learning
\item Experiential learning (hands-on learning \& physicalization)
\item Visualizations for public education (e.g., health education, science communication)
\item Engagement with visualizations
\item Teaching approaches that encourage creativity and design critique
\item Accessibility of visualization learning resources
\end{itemize}

\section{Submission Formats}
\label{submissionformats}
The workshop will accept three types of submissions, peer-reviewed by at least two PC members and one workshop organizer.
\begin{itemize}
        \setlength{\itemsep}{2pt}
        \setlength{\parskip}{1pt}
        \setlength{\parsep}{0pt}
    \item \textbf{Paper submissions:} full papers (6-8 pages excluding references, intended for publication through the IEEE DL) and short papers (2-4 pages excluding references, not for publication in IEEE DL) 
    \item \textbf{Activities:} practical reports that describe how an actitivity is conducted and how it could be reused by others and in other contexts; these activity reports will be published on the workshop website. 
    \item \textbf{Practical reports (curated):} experiences and reflections on teaching and education experiences that serve the exchange between newly appointed and experienced faculty.
\end{itemize}

\section{Workshop Activities}
\label{workshop_activities}
We plan a full-day workshop that will include an invited keynote, the presentation of the paper submissions, a hands-on session with datavis activities, a working group session, and a general discussion on community building and sustainability of the workshop. \\
\textbf{Keynote speakers:} We are in close contact with Dr.~Roberto Martinez-Maldonado and Prof.~Dragan Gavsevic from the Center for Learning Analytics at Monash University (Melbourne). Both are world-wide experts in the area of education for visualization and learning analytics and have previously worked with one of the workshop organizers. \\
\textbf{Submission presentation:} Each accepted paper will be given 5-10min to present, followed by a Q\&A session. The timing will depend on the submission type. After all accepted submissions have been presented, we will leave space for an open discussion on challenges and directions related to the workshop topic. \\
\textbf{Hands-on session:} We plan to invite the workshop participants to engage in and experience some selected data visualization activities actively. Workshop organizers will select these activities beforehand (e.g., from submissions and previously collected activities) and carry them out in parallel in small groups. \\
\textbf{Working groups:} Based on the discussions and experiences gathered during the activities in the first part of the workshop, we will collect questions and challenges to form working groups for the second part of the workshop. These working groups are formed around high-level questions and challenges regarding visualization education, literacy, and activities. Each working group is moderated by one of the workshop organizers and is asked to debate a given challenge acting like a task force that proposes potential solutions and concrete steps. Subsequently, each group will report their discussion for not longer than 5 minutes to the entire workshop.\\ 
\textbf{Steps forward:} To end the workshop, we will convene a structured discussion about ways forward for research in the area of visualization education, literacy, and activities, as well as in building a sustainable community in these areas. 
After the workshop, a voluntary workshop dinner will be planned to encourage community building.

\section{Tentative Schedule}
We are applying for a full-day workshop to be held on-site. The workshop will require a standard conference session room that can fit 50 people, with sound, visual equipment, internet access and preferably with large tables and chairs that can be moved around.

\begin{itemize}
\setlength{\itemsep}{2pt}
  \setlength{\parskip}{1pt}
  \setlength{\parsep}{0pt}

\item 09:00 \textemdash 09:15 \textbf{Opening and outline}
\item 09:15 \textemdash 10:15 \textbf{Keynote}
\textit{\item 10:15 \textemdash 10:45 \textbf{Coffee Break}}
\item 10:45 \textemdash 12:00 \textbf{Submission presentations and discussion}
\textit{\item 12:00 \textemdash 14:00 \textbf{Lunch Break}}
\item 14:00 \textemdash 15:15 \textbf{Hands-on session}
\textit{\item 15:15 \textemdash 15:45 \textbf{Coffee Break}}
\item 15:45 \textemdash 16:30 \textbf{Working groups}
\item 16:30 \textemdash 17:00 \textbf{Wrap-up discussion \& next steps}
\item 19:00 \textbf{Voluntary Workshop Dinner}
\end{itemize}

If given a half-day workshop, we plan on adjusting the schedule accordingly by excluding the keynote and planning a shorter hands-on session. 

\section{Workshop Organization Timeline}
The timeline for the workshop organization is as follows:
\begin{itemize}
    \setlength{\itemsep}{2pt}
    \setlength{\parskip}{1pt}
    \setlength{\parsep}{0pt}
\item March 20, 2023: \textbf{Call for Participation} 
\item July 1, 2023: \textbf{Submission Deadline}
\item July 25, 2023: \textbf{Reviews Collected}
\item July 30, 2023: \textbf{Author Notification}
\item August 15, 2023: \textbf{Submission Camera Ready Deadline}
\item September 20, 2023: \textbf{Deadline for Activity Submission}
\end{itemize}

We plan to advertise on the respective mailing lists for ACM CHI, IEEE VIS, DRS, ACM DIS, Digital Humanities, Art+Design, Tableau, and social media (Twitter, LinkedIn, etc.).

\section{Intended Outcomes}
This workshop will allow participants to discuss the challenges they face in data visualization education and exchange ideas and approaches with other visualization researchers and educators. Participants will sketch a research agenda and highlight the opportunities for visualization education, literacy, and activities for the community. This will contribute to the broader visualization community, literacy, and education agenda. We will collect activities and teaching material through the submission of papers and practicals. We plan to develop a website summarizing the collected activities and pointers to external teaching material to support newly appointed faculty in finding appropriate data visualization activities for the intended audiences and goals. 

\section{Program Committee}
\label{pccommittee}

\begin{enumerate}
\setlength{\itemsep}{2pt}
  \setlength{\parskip}{1pt}
  \setlength{\parsep}{0pt}
  \item Wolfgang Aigner (University of Applied Sciences St. Pölten)
  \item Jan Aerts (Amador Bioscience – Hasselt, Hasselt University \& KU Leuven)
  \item Lyn Bartram (Simon Fraser University)
  \item Enrico Bertini (NYU Tandon School of Engineering)
  \item Rahul Bhargava (Northeastern University)
  \item Magdalena Boucher (University of Applied Sciences St. Pölten)
  \item Alexandra Diehl (Universität Zürich)
  \item Marian Dörk (University of Applied Sciences Potsdam)
  \item Jason Dykes (City – University of London)
  \item Yuri Engelhardt (University of Twente)
  \item Kyle Hall (TD)
  \item Uta Hinrichs (University of Edinburgh)
  \item Xavier Ho (Monash University)
  \item Dietrich Kammer (University of Applied Sciences Dresden)
  \item Doris Kosminsky (University of Rio de Janeiro)
  \item Søren Knudsen (IT University of Copenhagen)
  \item Robert S Laramee (University of Nottingham)
  \item Tatiana Losev (Simon Fraser University – Burnaby)
  \item Areti Manataki (University of St Andrews)
  \item Isabel Meirelles (The Ontario College of Art and Design University)
  \item Luiz Morais (INRIA – Bordeaux)
  \item Till Nagel (Hochschule Mannheim)
  \item Arran Ridley (Independent researcher)
  \item Panagiotis Ritsos (Bangor University)
  \item Jon Schwabish (Urban Institute)
  \item Yalong Yang (Department of Computer Science at Virginia Tech)
  \item Wesley Willett (University of Calgary)
\end{enumerate}

\section{Organizing Committee}

\noindent\textbf{Mandy Keck}, mandy.keck@fh-hagenberg.at
\newline (\url{https://pure.fh-ooe.at/en/persons/mandy-keck})
\newline
Mandy Keck is a professor in UX and Interaction Design at the University of Applied Sciences Upper Austria, Campus Hagenberg, Austria. Her research focuses on data visualization literacy, visual exploration and recommender systems, and interaction design. Mandy is responsible for eight courses on data visualization at the University of Applied Sciences Upper Austria that target  different levels and audiences. She designed and conducted several hands-on workshops dealing with the creation of interface metaphors \cite{keck2014} and information visualizations \cite{keck2021} and was co-organizer of several conference workshops, including the IEEE VIS Datavis Activities workshops in 2020 and 2021 \cite{huron2020ieee, huron2021ieee}.\\ 

\noindent\textbf{Samuel Huron}, samuel.huron@cybunk.com
\newline (\url{https://www.telecom-paris.fr/samuel-huron})
\newline
Samuel Huron is an associate professor in Design and ICT at Telecom Paris Tech. His research focuses on creating and studying new tools to democratize dynamic information visualization authoring and by studying design methods apply to research. For his work on ``Constructive Visualization'' he received the 2015 best doctoral dissertation award from IEEE VGTC Pioneer Group. He designed and conducted many different workshops focus on visualization and physicalization during the last 5 years~\cite{huron2017let,VisKit}. 
Before, he was the lead designer of the Institute of Research and Innovation of the Pompidou Center.\\

\noindent\textbf{Georgia Panagiotidou}, g.panagiotidou@ucl.ac.uk
\newline{(\url{https://uclic.ucl.ac.uk/people/georgia-panagiotidou})
\newline
Georgia Panagiotidou is a Post-Doctoral researcher at UCL Interaction Centre in London, UK. Her work focuses on how to make visualization more inclusive by understanding how people handle data issues such as biases, uncertainties and frictions. She approaches data visualization as both a process and an outcome and has prepared and deployed data visualization activities using storytelling, gamification~\cite{Panagiotidou2020b} and physicalisation~\cite{Panagiotidou2020} among others.
\newline

\noindent\textbf{Christina Stoiber}, cstoiber@fhstp.ac.at
\newline \url{https://icmt.fhstp.ac.at/en/team/christina-stoiber}
\newline
Christina is a researcher and lecturer at the St. Pölten University of Applied Sciences, Austria. Her research interests are Information Visualization, HCI, Usability, Visualization Education, and Visualization Literacy. In October 2017, she started her dissertation on visualization literacy and visualization onboarding concepts for complex visualization tools, which she will finish in the summer of 2023. Therefore, one of her priorities is the integration of visualization onboarding concepts to support different domains in using complex visualization and visual analytics tools. She is also a certified examiner (SystemCert) for the UX-Development certificate based on ISO 17024.
\newline

\noindent\textbf{Fateme Rajabiyazdi}, fateme.rajabiyazdi@carleton.ca
\newline (\url{http://healthvisfutures.sce.carleton.ca/})
\newline
Fateme Rajabiyazdi is an Assistant Professor in the
Department of Systems and Computer Engineering
at Carleton University. Before joining Carleton, she
was a postdoctoral researcher at McGill University
Health Center, where she was a recipient of Fonds de la Recherche en Santé du Quebec postdoc scholarship 2020. She received her Ph.D. in Computer Science in the area of information visualization from the
University of Calgary in 2018. Her
research interests include visualizations that support
patient-healthcare provider communication
\newline

\noindent\textbf{Charles Perin}, cperin@uvic.ca
\newline (\url{http://charlesperin.net/})
\newline
Charles Perin is an Assistant Professor of Computer Science at the University of Victoria.
His research focuses on designing and studying new interactions for visualizations and on understanding how people may make use of and interact with visualizations in their everyday lives, including mobile and physical visualization. He teaches the visualization course at the University of Victoria in a student-centered, activity-based, flipped classroom format. He has co-organized five workshops at VIS (2015, 2017, 2020, 2021, 2022). He has presented work at the VIS workshop on pedagogy of data visualization~\cite{eggermont:biomimicry} and has published a reflection of using personal data physicalization in a visualization course~\cite{perin:2021:personal-data-phys}.
\newline

\noindent\textbf{Jonathan C. Roberts}, j.c.roberts@bangor.ac.uk
\newline (\href{http://www.bangor.ac.uk/computer-science-and-electronic-engineering/staff/jonathan-roberts/en}{[click-for-homepage]})
\newline
Jonathan is a professor in Visualization at Bangor University. He is the creator of the Five Design-Sheet method~\cite{roberts2015sketching} and lead author of the book Five Design-Sheets: Creative Design and Sketching for Computing and Visualization, Springer Nature, June 2017.  His research spans heritage, archaeology, oceanography, pedagogy, lexicography, and social networking domains, and for many years has encouraged researchers to develop multiple coordinated view systems.
He is a keen advocate of sketching and low-fidelity design~\cite{roberts2015sketching,CriticalThinkingSheets2020}, and promotes more design thinking in teaching.\newline

\noindent\textbf{Benjamin Bach}, bbach@ed.ac.uk 
\newline (\url{http://benjbach.net})
\newline
Benjamin is an Associate Prof. in Design Informatics and Visualization at the University of Edinburgh. His research designs and investigates interactive information visualization interfaces to help people explore, communicate, and understand data across media such as screens, mixed reality, paper, and physicalizations. Benjamin is involved in teaching four courses on data visualizations at the University of Edinburgh, targeting different audiences and levels, including one online course for professional development for which is co-leading (together with Uta Hinrichs) funding of more than \pounds160k, including the creation of activities and material. In the past, Benjamin has been working on Cheatsheets~\cite{wang2020cheatsheets}, visualization tools~\cite{marketplace}, data comics~\cite{bach2018design}, and data comic workshops~\cite{wang2019teaching} for data visualization and is currently working on understanding users and support them in learning interactive tools.
\newline

\acknowledgments{
We would like to thank all participants of the Dagstuhl seminar ``22261" for the inspiring discussions during the seminar and their support in developing this workshop proposal.}

\bibliographystyle{abbrv-doi}

\bibliography{template}
\end{document}